\documentclass[preprint,showpacs,preprintnumbers,amsmath,amssymb,prd,superscriptaddress]{revtex4}  

\usepackage{graphicx}% Include figure files
\usepackage{dcolumn}% Align table columns on decimal point
\usepackage{bm}% bold math

\newcommand{\Apj}{Astrophys. J. }

\renewcommand{\figurename}{Fig.}
\renewcommand{\tablename}{Table.}

\begin{document}

\title{
Observational Constraint on Heavy Element Production in 
Inhomogeneous Big Bang Nucleosynthesis
}

\author{Riou Nakamura
\footnote{E-mail: riou@phys.kyush-u.ac.jp}}
\affiliation{Department of Physics, Faculty of Sciences, Kyushu University,
%6-10-1 Hakozaki, Higashi-ku, 
Fukuoka 812-8581, Japan}

\author{Masa-aki Hashimoto}
\affiliation{Department of Physics, Faculty of Sciences, Kyushu University,
%6-10-1 Hakozaki, Higashi-ku, 
Fukuoka 812-8581, Japan}

\author{Shin-ichiro Fujimoto}
\affiliation{Department of Control and Information Systems Engineering, 
Kumamoto National College of Technology, Kumamoto 861-1102, Japan}

\author{Nobuya Nishimura}
\affiliation{National Astronomical Observatory, 
2-21-1, Osawa, Mitaka, Tokyo 181-8588, Japan}

\author{Katsuhiko Sato}
\affiliation{
Institute for the Physics and Mathematics of the Universe,
The University of Tokyo, 
%5-1-5 Kashiwanoha, Kashiwa, 
Chiba, 277-8568, Japan
%Department of Physics, School of Science, University of Tokyo,
%7-3-1 Hongo, Bunkyo, Tokyo 113-0033, Japan
}
\affiliation{
School of Science and Engineering, Meisei University, 
Tokyo 191-8506, Japan
}

\begin{abstract}
Based on a scenario of the inhomogeneous big-bang nucleosynthesis (IBBN),
 we  investigate the detailed nucleosynthesis that includes the production
 of  heavy
 elements beyond ${}^7_{}$Li. From the observational constraints on
 light elements of ${}^4_{}$He and D for the baryon-to-photon ratio given by
WMAP, possible regions 
 found on the plane of the volume fraction of the high density region
 against the ratio between high- and low-density regions. 

 In these allowed regions, we have confirmed that the heavy elements
 beyond Fe can be produced appreciably, where $p$- and/or $r$-process elements are
 produced well simultaneously compared to the solar system abundances. We suggest that
recent observational signals such as ${}^4_{}$He
 overabundance in
 globular clusters and high metallicity abundances in quasars
could be partly due to the results of IBBN. Possible implications are given for the
formation of the first generation stars.
\end{abstract}

\date{\today}

\pacs{26.35.+c, 98.80.Ft, 13.60.Rj}
\maketitle

\section{INTRODUCTION}
\label{sec:ibbn_intro}

Big bang nucleosynthesis  has been investigated mainly on the context of
the standard cosmological model (SBBN), where origin of light elements
of $^{4}$He, D, and $^{7}$Li have been discussed in detail~\cite{Iocco:2008va}.  
While  observations of ${}^{4}_{}$He are
still in debate with the uncertainty of 20-30 \% 
in the abundance~\cite{Luridiana2003,OliveSkillman04,Izotov:2007ed},
those of D constrain severely the possible range of the abundance
production in the early universe~\cite{Kirkman2003,OMeara2006,Pettini2008}. 
Contrary to the above standard BBN, the heavy element nucleosynthesis
beyond the mass number $A=8$ has been proposed from twenty years 
ago~\cite{IBBN0,IBBN1,TerasawaSato89,Alcock1987,2zone,Jedamzik1994,Matsuura:2004ss}, 
where the model is called the inhomogeneous BBN (IBBN). 
This model relays on the inhomogeneity of baryon concentrations that could
be induced by baryogenesis~(e.g. Ref.~\cite{Matsuura:2004ss}) or some
phase transitions such as QCD or electro-weak phase
transition~\cite{Alcock1987,Fuller1988,IBBN_QCD} during the expansion of the
universe. Although a large scale inhomogeneity is inhibited by many 
observations~\cite{WMAP3,WMAP5}, small scale one has been advocated within the present
accuracy of the observations. Therefore, it remains a possibility for
IBBN to occur in some degree during the early era.
 
 On the other hand, Wilkinson Microwave Anisotropy Probe (WMAP) has derived
critical parameters concerning the cosmology of which the present 
baryon-to-photon ratio $\eta$ is determined to be 
$\eta = \left( 6.19\pm 0.15 \right)\times 10^{-10}_{}$~\cite{WMAP5}.
This value is almost consistent with that obtained from the observation
of D. Therefore, considering the uncertainty of the ${}^{4}_{}$He
abundance, we can fix the ratio $\eta$ in the discussion of the
nucleosynthesis in the early universe. If the present ratio of $\eta$ is
determined,  BBN can be performed along that line in the thermodynamical history with
use of the nuclear reaction network. 
On the other hand, peculiar observations of abundances for
heavy elements and/or ${}^{4}_{}$He could be understood in the way of IBBN.
For example, the quasar metallicity of C, N, and Si could have been explained
from IBBN~\cite{Juarez2009}.
Furthermore, from recent observations of globular clusters, possibility
of inhomogeneous helium distribution is pointed out~\cite{Moriya2010},
where some separate groups of different main sequences in blue band of low mass stars are
assumed due to high primordial helium abundances compared to the
standard value \cite{Bedin2004,Piotto2007}.

Despite a negative opinion against IBBN due to insufficient consideration
of the scale of the inhomogeneity~\cite{Rauther2006},
Matsuura et al. have found that the heavy element synthesis for both $p$- and 
$r$-processes is possible if $\eta > 10^{-4}$ \cite{Matsuura2005}, where
they have also shown that the high
$\eta$ regions are compatible with the observations of the light elements,
$^4$He and D~\cite{Matsuura2007}.
However, their analysis is only limited to a parameter of
a specific baryon number concentration. 
Therefore, it should be needed to constrain the possible
regions from available observations in the wide parameter space that describes
 the IBBN.

In \S \ref{sec:Numerical}, we review and give the adopted model of
IBBN~\cite{Matsuura2007}. Constraints on the critical parameters of IBBN
due to light element observations are shown in \S III,  and the
productions of possible heavy element nucleosynthesis is presented in \S IV.
Finally, \S \ref{rec:summary} is devoted to the summary and discussion.

\section{Cosmological Model}
\label{sec:Numerical}

We adopt the two-zone model for the inhomogeneous BBN,
where the early universe is assumed to have the high- and low- baryon
density regions~\cite{IBBN1} under the background temperature $T$.
For simplicity we ignore the diffusion effects before 
$\left( 10^{10}_{} {\rm K} < T < 10^{11}_{} {\rm K} \right)$ and
during the primordial nucleosynthesis $\left( 10^{7}_{} {\rm K} < T < 10^{10}_{} {\rm K} \right)$,
where the plausibility will be discussed in \S \ref{rec:summary}.
After the epoch of BBN, all the elements are assumed to be mixed homogeneously.

Let us define the notations,  $n^{}_{ave}, n^{}_{high}$, and $n^{}_{low}$ as
averaged-, high-, and low- baryon number densities. $f^{}_v$ is the volume
fraction of the high baryon density region. 
$X^{ave}_{i}, X^{high}_i$ and $X^{low}_{i}$
are mass fractions of each element $i$ in averaged-, high- and
low-density regions, respectively, 
Then, basic relations are written as follows: 
\begin{eqnarray}
n^{}_{ave} &=& f^{}_{v}n^{}_{high}+\left( 1-f^{}_v \right)n^{}_{low},
 \label{eq:num_b} \\
 n^{}_{ave}X^{ave}_{i} &=&  f^{}_{v}n^{}_{high} X^{high}_i 
  +\left( 1-f^{}_v \right)n^{}_{low} X^{low}_{i}. \label{eq:massfrac}
\end{eqnarray}
Here we assume the baryon fluctuation to be isothermal as was done in
previous studies~(e.g., Refs.~\cite{TerasawaSato89,Alcock1987,Fuller1988}).
Under that assumption, since the baryon-to-photon ratio is
defined by the number density of photon in standard BBN, 
$n^{}_\gamma=2\zeta{(3)}/\pi^2\left( k_BT/\hbar c\right)^3_{}$,
Eqs.\eqref{eq:num_b} and \eqref{eq:massfrac} are rewritten as follows:
\begin{eqnarray}
\eta^{}_{ave}  &=&
 f^{}_{v}\eta^{}_{high}+(1-f^{}_v)\eta^{}_{low},  \label{eq:eta_ave}
\\
\eta^{}_{ave}X^{ave}_{i}
 &=&
 f^{}_{v}X^{high}_{i}\eta^{}_{high}+(1-f^{}_{v})X^{low}_{i}\eta^{}_{low},
\label{eq:Yi_ave}
\end{eqnarray}
where $\eta$s with subscripts are the baryon-to-photon ratios in each
region. % and $n^{}_{high}/n^{}_{low}=\eta^{}_{high}/\eta^{}_{low}$.
In the present paper, we fix $\eta^{}_{ave}=6.1\times10^{-10}$ from the cosmic
microwave background observation~\cite{WMAP3,WMAP5}.
$\eta^{}_{high}$ and $\eta^{}_{low}$ are obtained from both $f^{}_v$ and
the density ratio between high- and low-density region:
 $R\equiv n^{}_{high}/n^{}_{low}=\eta_{high}/\eta_{low}$.

To calculate the evolution of the universe, 
we solve the following Friedmann equation,
\begin{equation}
\left(\frac{\dot{x}}{x}\right)^2 =\frac{8\pi G}{3}\rho , 
\end{equation}
where $x$ is the cosmic scale factor and $G$ is the gravitational constant.
The total energy  density $\rho$ is the sum of decomposed parts:
\[
\rho=\rho^{}_{\gamma}+\rho^{}_{e^{\pm}}+\rho^{}_{\nu}+\rho^{}_{b}.
\]
Here the subscripts $\gamma, e^{\pm}_{}, \nu$, and $b$ indicate 
photons, electrons/positrons, neutrinos, and baryons, respectively.
We note that $\rho^{}_b$ is the average value of baryon density obtained
from Eq.~(\ref{eq:num_b}).

The energy conservation law is used to get the time evolution of the
temperature and the baryon density,
\begin{equation}
\frac{d}{dt}(\rho x^3) + %\frac{p}{c^2}\frac{d}{dt}(x^3)
p\frac{d}{dt}(x^3)
=0,
\end{equation}
where $p$ is the pressure of the fluid.

\section{Constraints from light-element observations}
\label{Results}

In this section, we calculate the nucleosynthesis in high- and
low-density regions with use of the BBN code~\cite{Hashimoto1985} which
includes 24 nuclei from neutron to ${}^{16}_{}$O.
We adopt the reaction rates of NACRE~\cite{NACRE}, the neutron life time 
$\tau^{}_N = 885.7$ sec~\cite{Hagiwara:2002fs}, and take account of the
number of species of the massless neutrinos $N^{}_\nu=3$.

\begin{figure}[tb]
\begin{center}
 \includegraphics[width=0.8\linewidth,keepaspectratio]{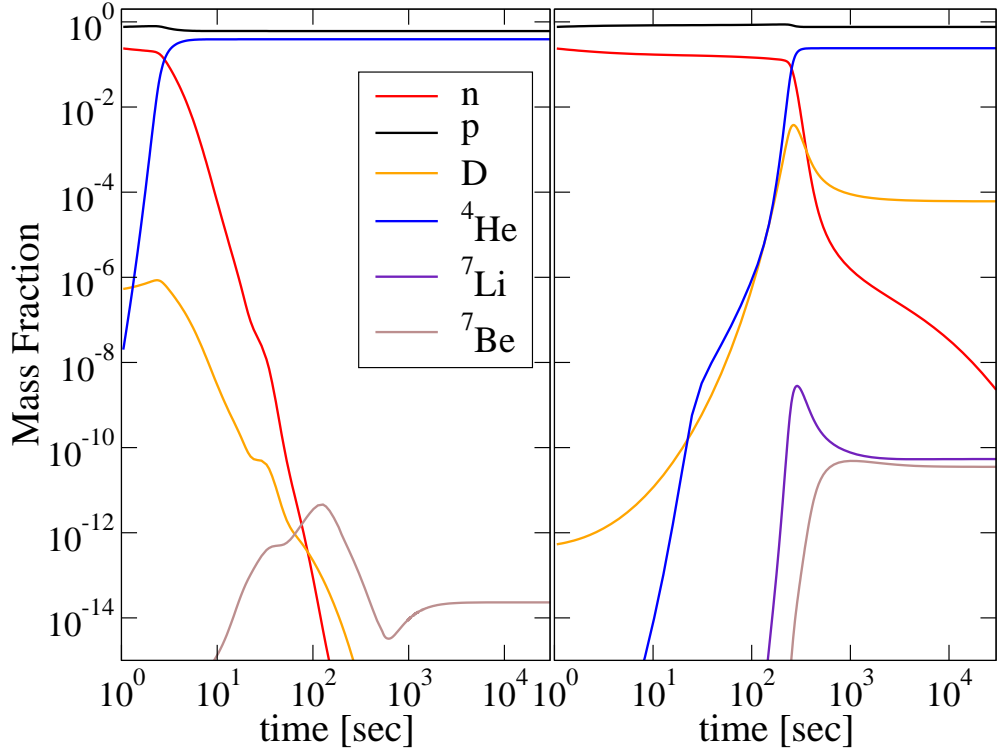}
 \caption{Illustration of the nucleosynthesis in the two-zone
 IBBN model with $f^{}_v=10^{-6}_{}$ and $R=10^{6}_{}$. 
 The baryon-to-photon ratios in the high- (left panel) and low- (right
 panel) density regions are  $\eta^{}_{high}=3.05\times10^{-4}$ and
 $\eta^{}_{low}=3.05\times 10^{-10}$,  respectively.}
\label{fig:2zone_bbn}
\end{center}
\end{figure}

Figure~\ref{fig:2zone_bbn} illustrates the light element synthesis
in the high- and low-density regions with $f_v=10^{-6}_{}$ and
$R=10^6_{}$ that correspond to $\eta^{}_{high}=3.05\times10^{-4}$ and
$\eta^{}_{low}=3.05\times10^{-10}$.
In the low-density region the evolution of the elements is almost the same as that
of standard BBN. In the high-density region, 
while ${}^4_{}$He  is more abundant than that in the low-density region,
${}^7_{}$Li (or ${}^7_{}$Be) is much less produced.
It implies that heavier nuclei such as ${}^{16}$O, hardly synthesized in SBBN,
are synthesized at high-density region.

For $f^{}_v\ll0.1$, the heavier elements can be synthesized in the
high-density regions as discussed in Ref.~\cite{Jedamzik1994}.
For $f^{}_v>0.1$, contribution of the low-density
region to $\eta^{}_{ave}$ can be neglected and therefore
to be consistent with observations of light elements, we need to
impose the condition of $f^{}_v<0.1$.
Now, we put constraints on $f^{}_v$ and $R$ by comparing the average
values of ${}^4_{}$He and D obtained from Eq.~(\ref{eq:Yi_ave}) with 
the following observational values. 
First we adopt the primordial ${}^4_{}$He abundance reported in
Ref.~\cite{OliveSkillman04}: 
\begin{equation}
0.232<Y^{}_p<0.258. \label{eq:Heobs}
\end{equation}
Next, we take the primordial abundance  from the
 D/H observation reported in Ref.~\cite{PDG2008}
\begin{equation}
\text{D/H}  =\left( 2.84\pm0.26 \right)\times 10^{-5}, \label{eq:Dobs}
\end{equation}
where the systematic error given in Ref.~\cite{OMeara2006} is adopted.

\begin{figure}[tb]
\begin{center}
 \includegraphics[width=0.8\linewidth,keepaspectratio]{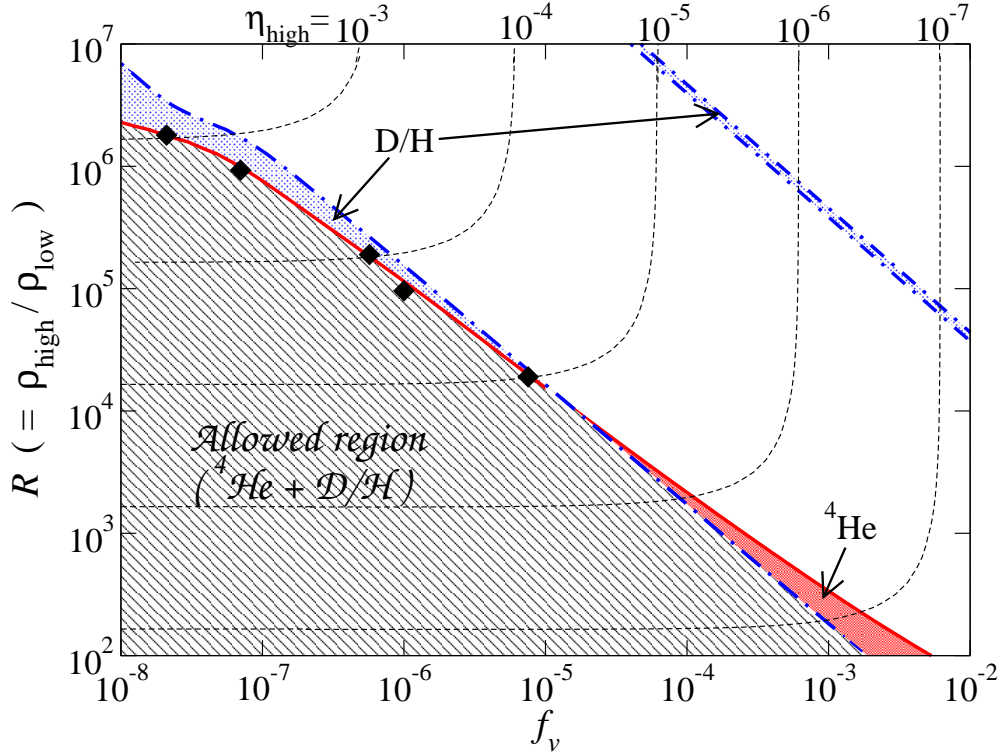}
\caption{Constraints on the $f_v-R$ plane from the observations of light
 element abundances. The region below the solid line is allowed one obtained
 from ${}^4_{}$He observation~\cite{OliveSkillman04}. 
 Constraints from the D/H observation~\cite{PDG2008} are shown by the
 region below the dot-dashed line. The shaded region is the allowed parameters determined
 from the two observations of ${}^4_{}$He  and D/H. The dotted lines
 show 
 the contours  of the baryon-to-photon ratio in the high-density
 region. Filled squares indicate the parameters for heavy element
 nucleosynthesis adopted in \S \ref{sec:Result2}.}
\label{fig:HeDcntr}
\end{center}
\end{figure}

Figure~\ref{fig:HeDcntr} illustrates the constraints on the $f^{}_v-R$ plane
from the above light-element observations with contours of constant $\eta^{}_{high}$.
The solid and dashed lines indicate the upper limits from Eqs.~(\ref{eq:Heobs})
and (\ref{eq:Dobs}), respectively.
As the results, we can obtain approximately the following relations between $f^{}_v$ and $R$ :
\begin{equation}
 R \leq 
\begin{cases}
0.26\times f_v^{-0.96} & \text{~for~} f_v>3.2\times10^{-6}, \\
1.20\times f_v^{-0.83} & \text{~for~} f_v\leq 3.2\times10^{-6} .
\end{cases}
\label{eq:HeDlimit_} 
\end{equation}
As shown in Figure~\ref{fig:HeDcntr}, we can find the allowed regions
which include 
the very high-density region such as $\eta^{}_{high}=10^{-3}$. 

Matsuura et al.~\cite{Matsuura2007} defined a parameter of the baryon number concentration
$a$ in the high density
region instead of two parameters of $f^{}_v$ and $R$ that are needed to
solve Eqs.\eqref{eq:eta_ave} and \eqref{eq:Yi_ave}:
\[
 f^{}_v\eta^{}_{high}: \left( 1-f^{}_v\right)\eta^{}_{low}
 = a : \left( 1-a \right) .
\]
However, they have only examined the case of $\eta^{}_{high}=10^{-3}$ and
$\eta^{}_{low}=3.162\times10^{-10}$, where $a=0.48$ for $\eta_{ave}=6.1\times10^{-10}_{}$. 
Our constraints in Eq.~(\ref{eq:HeDlimit_}) correspond to $a=0.02-0.5$.
Since we have fixed the value of $\eta_{ave}$, we can obtain physically more reasonable
regions on the plane of $(f^{}_v, R)$.

\begin{figure}[tb]
\begin{center}
  \includegraphics[width=0.80\linewidth,keepaspectratio]{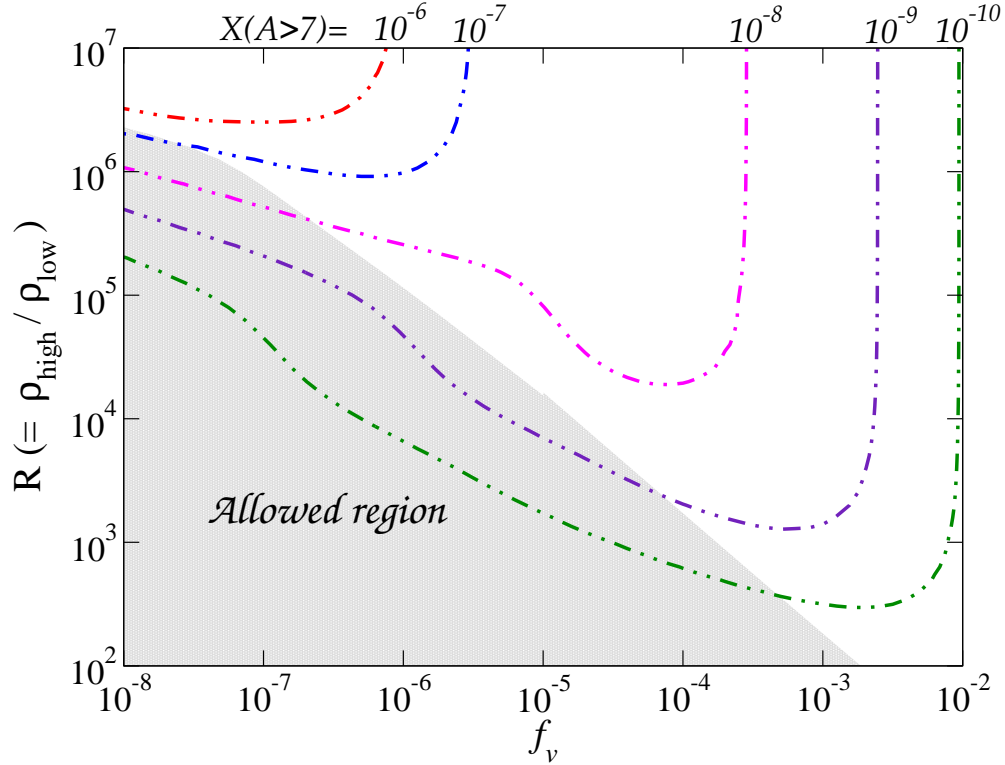}
\caption{Contours of the averaged total mass fractions which are the sum
 of nuclei heavier than  ${}^7_{}$Li, where we find consistent regions with ${}^4_{}$He
 and D observations.}
\label{fig:cntr_upLi7}
\end{center}
\end{figure}

Naturally, as $\eta^{}_{high}$ takes larger value, nuclei which are heavier
than ${}^7_{}$Li are synthesized  more and more. Then we can estimate
the amount of total CNO elements in the allowed region.
Figure~\ref{fig:cntr_upLi7} illustrates the contours of the summation of
the average values of the heavier nuclei~($A>7$), which correspond to
Fig.~\ref{fig:HeDcntr} and are drawn using the constraint from
${}^4_{}$He and D/H observations .
As a consequence, we get the upper limit of total mass 
fractions for heavier nuclei as follows: $X(A>7) \leq 10^{-7}_{}$.

We should note that abundance flows proceed beyond the CNO elements
thanks to the larger network for high
$\eta$-values as shown in Table~\ref{tab:abundance_4463_2} of the following section.

\section{Heavy element Production}
\label{sec:Result2}

In the previous section, we have obtained the amount of CNO elements produced
in the two-zone IBBN model.
However, it is not enough to examine the nuclear production beyond $A>8$ because the
baryon density in the high-density region becomes so high that elements
beyond CNO isotopes can be produced~\cite{Wagoner1967,2zone,Matsuura:2004ss,Matsuura2005}.

In this section, we investigate the heavy element nucleosynthesis in the
high-density region considering the constraints shown in Fig.~\ref{fig:HeDcntr}.
The temperature and density evolutions are the same as used in the previous section.
Abundance change is calculated with a large nuclear reaction
network, which includes 4463 nuclei from neutron $(n)$, proton $(p)$ to Americium 
({\it Z} = 95 and {\it A} = 292).
Nuclear data, such as reaction rates, nuclear masses, and partition
functions, are the same as used in~\cite{fujimoto} except for the
neutron-proton interaction; We take the weak interaction rates between n and p
%$n\leftrightarrow p$ reaction rates
from Kawano code~\cite{Kawano}, which is adequate for the high temperature
epoch of $T>10^{10}$~K.
We note that mass fraction of ${}^4_{}$He and D obtained with the large
network are consistent with those in in \S \ref{Results} within the
accuracy of few percents.

As seen in Fig.~\ref{fig:cntr_upLi7}, heavy elements of
$X(A>7)>10^{-9}$ are produced nearly along the upper limit of $R$.
Therefore, to examine the efficiency of the heavy element production, 
we select five models with the following parameters:
$\eta^{}_{high}=10^{-3}_{}, 5.3\times10^{-4}, 10^{-4}_{}, 5.5\times10^{-5}$
, and $10^{-5}$ corresponded to
$\left( f^{}_v, R \right)= \left( 2.1\times10^{-8}, 1.8\times10^{6}_{} \right)$
, $\left( 7.0\times10^{-8}, 9.3\times10^{5}_{} \right)$
, $\left( 5.7\times10^{-7}, 1.9\times10^{5}_{} \right)$
, $\left( 1.0\times10^{-6}, 9.6\times10^{4}_{} \right)$
, and $\left( 7.5\times10^{-6}, 1.9\times10^{4}_{} \right)$.
Adopted parameters are indicated by filled squares in Fig.~\ref{fig:HeDcntr}.

\begin{figure}[t]
\begin{center}
 \includegraphics[width=0.75\linewidth,keepaspectratio]{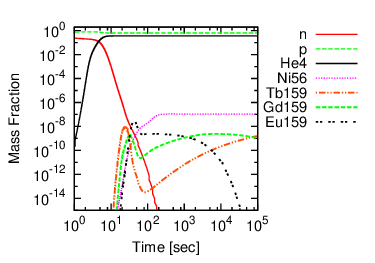}

 {(a) $\eta^{}_{high}=1.02\times10^{-4}_{}$}

 \includegraphics[width=0.75\linewidth,keepaspectratio]{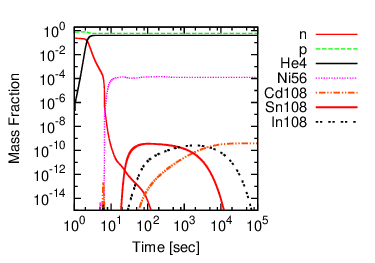}

 {(b) $\eta^{}_{high}=1.06\times10^{-3}_{}$}
\end{center}
\caption{Time evolution of the mass fractions in high-density regions of
  (a) $\eta^{}_{high}=1.02\times10^{-4}_{}$ and (b) $\eta^{}_{high}=1.06\times10^{-3}_{}$. }
\label{fig:MFNHSfig4}
\end{figure}

Figure~\ref{fig:MFNHSfig4} shows the results of nucleosynthesis in the
 high-density regions with $\eta^{}_{high}\simeq 10^{-4}_{}$ and $10^{-3}$.
 For $\eta^{}_{high}\simeq 10^{-4}$, the nucleosynthesis paths are
 classified with the mass number~\cite{Matsuura2005}.
 For nuclei of mass number $A \leq 100$, proton captures are
 very active  compared to the neutron capture of $T>2\times10^{9}$~K and
 the path moves to the proton rich side, which began by breaking out of
 the hot CNO cycle.
 For nuclei of $100 < A < 120$,
 the path goes across the stable nuclei from proton 
 to neutron rich side, since the temperature decreases and the number of
 seed nuclei of the neutron capture process increase significantly. 
Concerning heavier nuclei of ${\it A} \ge 120$, neutron
 captures become much more efficient. 
 In Figure~\ref{fig:MFNHSfig4}(a), we see the time evolution of the
 abundances of Gd and Eu for the mass number 159. 
First ${}^{159}_{}$Tb (stable $r$-element) is synthesized and later
 ${}^{159}_{}$Gd and ${}^{159}_{}$Eu are synthesized through the neutron
 captures. After $t=10^3_{}$ sec, ${}^{159}_{}$Eu decays to nuclei
 by way of ${}^{159}_{}$Eu $\rightarrow {}^{159}_{}$Gd $\rightarrow {}^{159}_{}$Tb,
where the lifetimes of ${}^{159}_{}$Eu and ${}^{159}_{}$Gd are $10.1$ min
 and $18.479$ h, respectively. These neutron capture process is not
 similar to the canonical $r-$process, since the nuclear processes
 proceed under the condition of the high-abundance of protons.

 For $\eta^{}_{high}\simeq 10^{-3}$, the reactions first proceed along
 the stable line, because triple-$\alpha$ reactions and other particle
 induced reactions are very effective.  Subsequently, the reactions directly proceeds to the
 proton rich region, through rapid proton captures.
 As shown in  Fig.~\ref{fig:MFNHSfig4}(b), ${}^{108}_{}$Sn which is
 proton-rich nuclei is synthesized. After that, stable nuclei
 ${}^{108}_{}$Cd is synthesized by way of
 ${}^{108}_{}$Sn $\rightarrow {}^{108}_{}$In $\rightarrow {}^{108}_{}$Cd,
 where the lifetimes of  ${}^{108}_{}$Sn and ${}^{108}_{}$In are 
$10.3$ min and  $58.0$ min, respectively.
 In addition, we notice  the production of radioactive nuclei of ${}^{56}_{}$Ni
and ${}^{57}_{}$Co, where
${}^{56}_{}$Ni is produced at early times, just after the formation of ${}^4_{}$He.
Usually,  nuclei such as ${}^{56}_{}$Ni and ${}^{57}_{}$Co %(it
				%decays into ${}^{56}_{}$Fe) 
are produced in supernova explosions, which are assumed to be the events
after the first star formation (e.g. Ref.~\cite{Hashimoto1995}). In IBBN
model, however, this production can be found to occur at extremely high density
region of $\eta^{}_{high}\geq 10^{-3}_{}$ as the primary elements
without supernova events in the early universe.

\begin{figure}[t]
\begin{center}
 \includegraphics[width=0.75\linewidth,keepaspectratio]{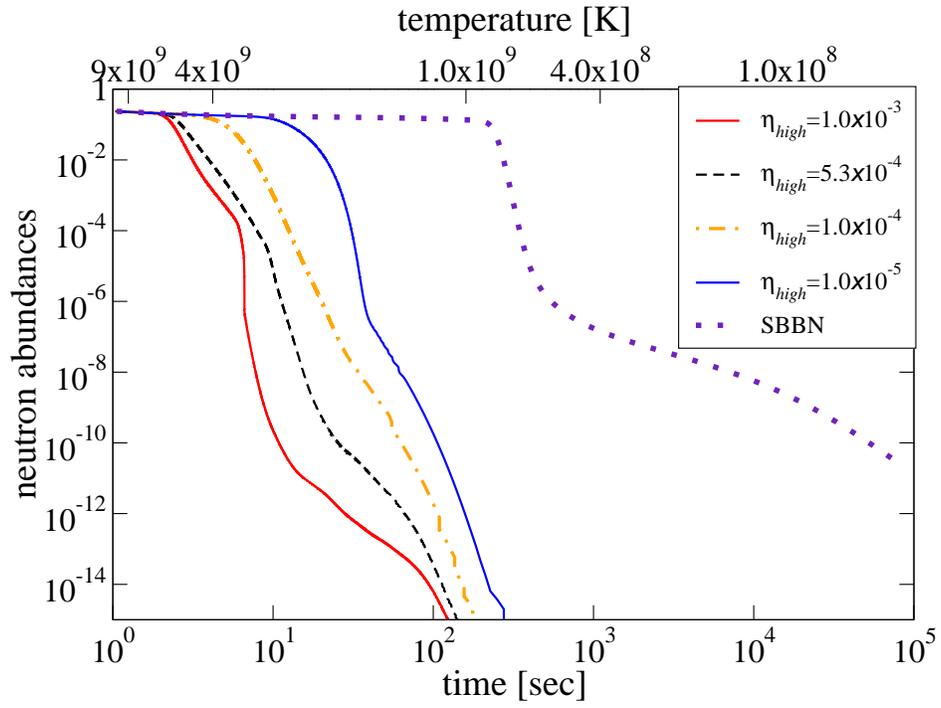}
\end{center}
\caption{\label{fig:n2p} Time evolution of the neutron abundance in SBBN
 ($\eta=6.1\times10^{-10}$) and IBBN in the high-density region.}
\end{figure}

To explain differences of the nuclear reactions which depend on the
baryon density, we focus on the neutron abundances.
Figure~\ref{fig:n2p} shows the evolutions of the neutron abundances in
the SBBN and IBBN models.
For $\eta_{high}=10^{-3}$, neutron abundance decreases rapidly at $\sim$ 10 sec
to the formation of ${}^{4}_{}$He and ${}^{56}_{}$Ni.
Thus, neutron abundance is not enough to induce the neutron capture
producing heavy nuclei of $A > 90$. On the other hand, neutron abundance tends to 
 remain even at the high temperature for the  lower value of $\eta^{}_{high}$.
We can see the case of $\eta^{}_{high}=10^{-4}$, where there remain much neutrons to occur the
neutron capture reaction.  
Thus the neutron capture process to produce heavy elements of $A > 90$ can become active.

\begin{figure}[t]
\begin{center}
 \includegraphics[width=0.9\linewidth,keepaspectratio,clip]{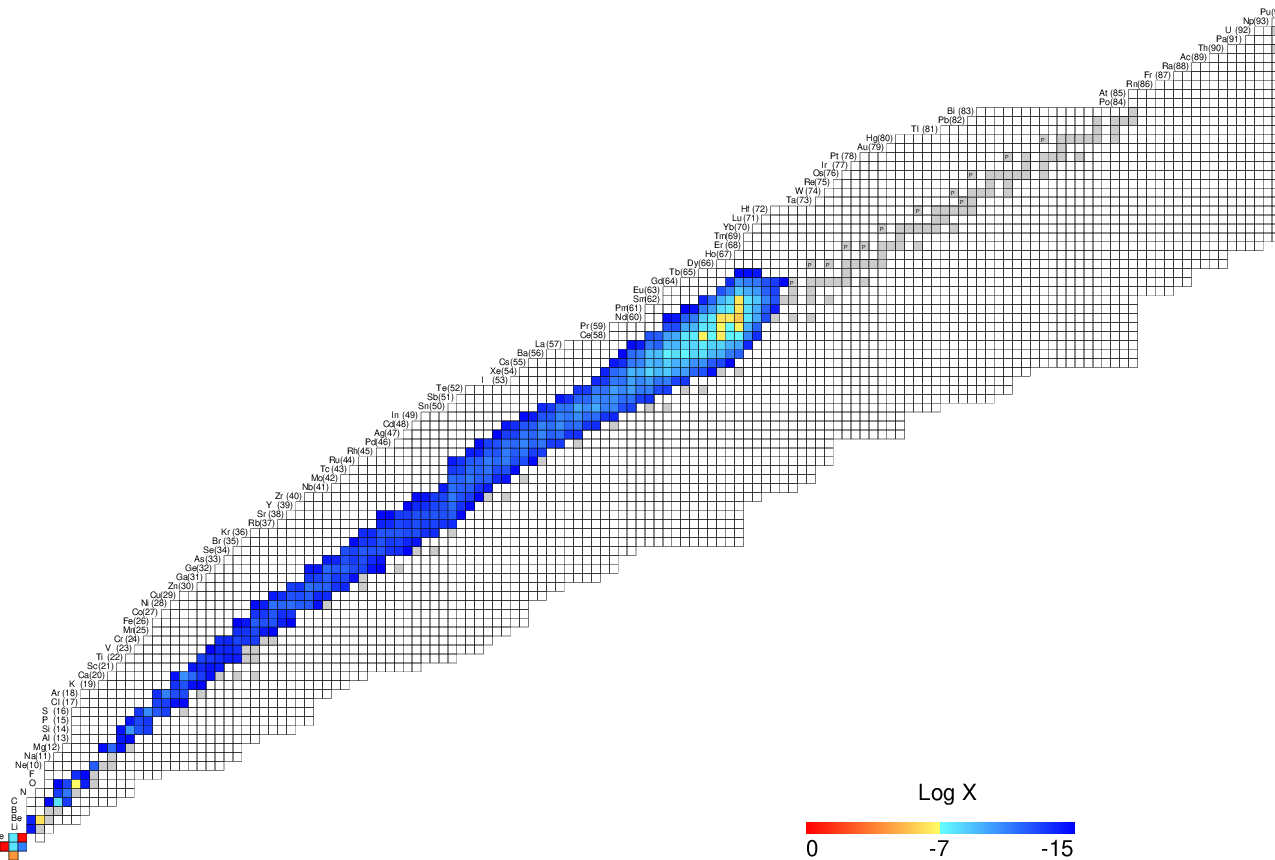}
\end{center}
\caption{\label{fig:flow1} Abundance distribution at
 $T=3.5\times10^{9}$ K ($t\sim9$ sec) in $\eta^{}_{high}=5.3\times10^{-4}$. The gray regions
 are those of stable nuclei. }
\end{figure}

\begin{figure}[t]
\begin{center}
 \includegraphics[width=0.9\linewidth,keepaspectratio,clip]{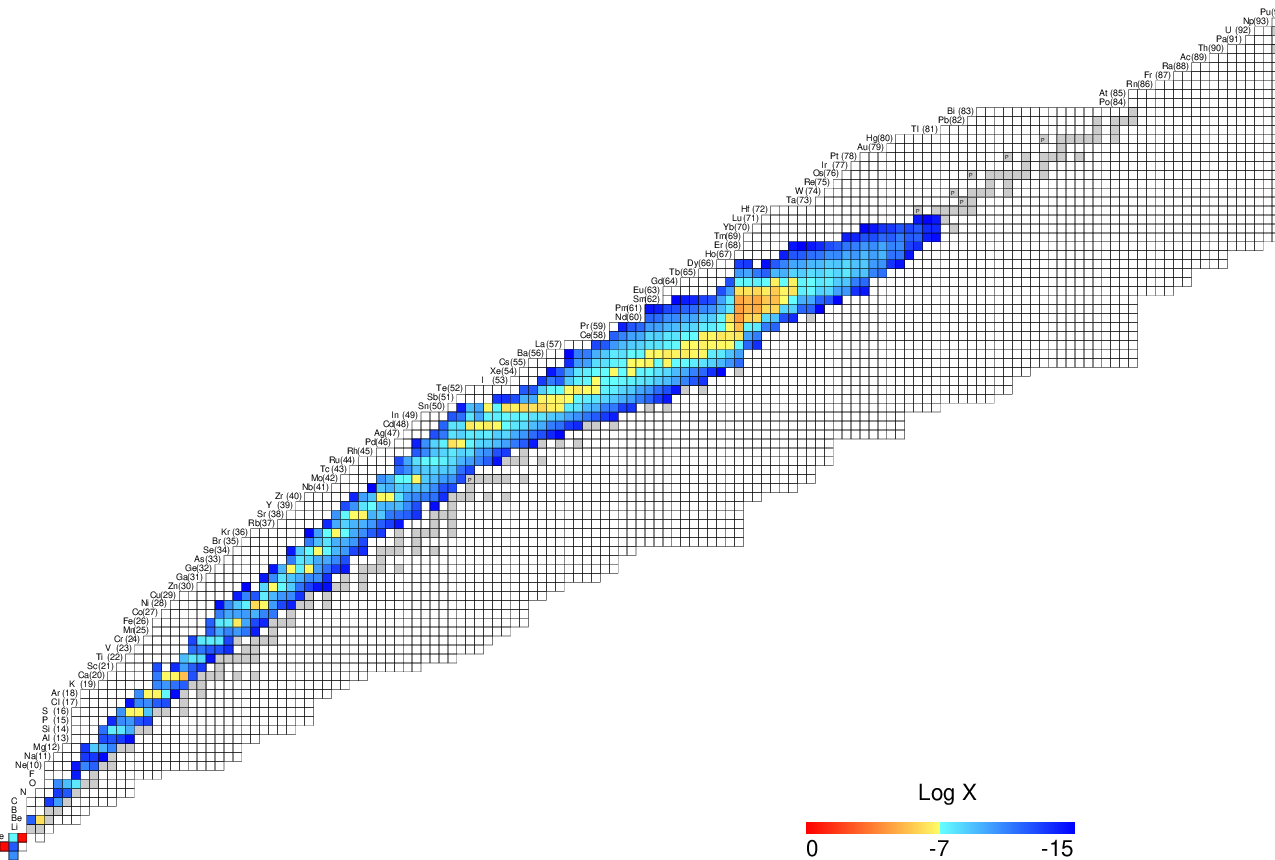}
\end{center}
\caption{\label{fig:flow2} . Abundances distribution at
 $T=1.9\times10^{9}$ K ($t\sim40$ sec) in $\eta^{}_{high}=5.3\times10^{-4}$.}
\end{figure}

 Time scales in the decrease for the neutron abundances change drastically the flow
 of the abundance production. Figures~\ref{fig:flow1} and \ref{fig:flow2} show the
 flows for $\eta^{}_{high}=5.3\times10^{-4}$. Before the significant
 decrease in the neutron abundances before $10$ sec, the nucleosynthesis
 proceeds already along the stable line by way of the neutron included
 reactions~(Fig.~\ref{fig:flow1}). 
At that time, the nuclear reactions are stuck around $Z=60$ with $N=82$,
 since it takes time to synthesize heavier nuclei because Nd ($Z=60$)
 and Sm ($Z=62$) have some stable isotopes. 
As time goes,  neutron captures of these nuclei start, where
the neutron captures proceed significantly and $r$-elements can be synthesized. 
After the  depletion of neutrons ($t>40$ sec), nuclei around the neutron numbers $N=82$ are
 produced through proton induced reactions such as ${}^{144}$Sm (Fig.~\ref{fig:flow2}).
% %---------------------------------------------------------------------------

Final results ($T=4\times10^{7}$~K) of nucleosynthesis calculations are shown in
Tables~\ref{tab:abundance_4463} and \ref{tab:abundance_4463_2}.
Table~\ref{tab:abundance_4463} shows the abundances of light elements, ${}^{4}_{}$He, D,
and ${}^{7}_{}$Li, in high- and low-density regions with their average values. 
Abundances of the low-density side (the third and sixth columns) are
obtained from the calculation by BBN code used in \S III, because
abundance flows beyond $A=7$ are negligible.
%The average values are consistent with the light-element observations.
We should note that the average abundances of ${}^{4}_{}$He and D are consistent with their
observational values of \eqref{eq:Heobs} and \eqref{eq:Dobs}.
Table~\ref{tab:abundance_4463_2} shows the amounts of heavy elements.
When we have calculate the average values, we set the
abundances of $A>16$ as zero for low-density side.
For $\eta^{}_{high}\simeq 10^{-4}_{}$, 
a lot of nuclei of $A>7$ are synthesized whose amounts are comparable
to that of ${}^{7}_{}$Li. Produced elements in this case include both
$s$-element (i.e. ${}^{138}_{}$Ba)
and $r$-elements (for instance, ${}^{142}$Ce and ${}^{148}$Nd), since
moderate amounts of neutrons remain as shown in Fig.~\ref{fig:n2p}

For $\eta^{}_{high}\simeq 10^{-3}_{}$, there are few $r$-elements while both
$s$-elements (i.e. ${}^{82}_{}$Kr and ${}^{89}_{}$Y) and $p$-elements (i.e ${}^{74}$Se and ${}^{78}$Kr) are synthesized such
as the case of supernova explosions.
Although heavy nuclei of $A \ge 100$ are not synthesized appreciably,
those of $A\leq~90$ are produced well owing to the explosive
nucleosynthesis under the high density 
circumstances ($\rho\sim10^6~\rm g~cm^{-3}$).
The most abundant element is found to be  ${}^{56}_{}$Ni
whose production value is much larger than the estimated upper limit of
the total mass fraction (shown in Fig.\ref{fig:cntr_upLi7}) derived from the BBN code calculations.
This is because our BBN code used in \S \ref{Results} includes the elements up to
$A=16$ and the actual abundance flow proceeds to much heavier elements.
%---------------------------------------------------------------------------

\begin{figure}[t]

\begin{center}
 \includegraphics[width=.70\linewidth,keepaspectratio]{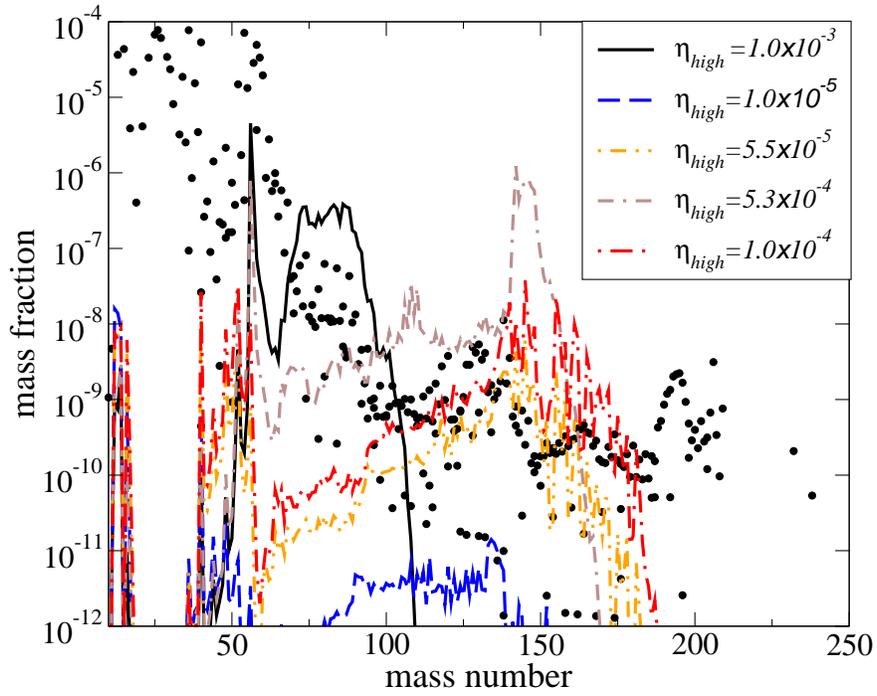}
\end{center}
\caption{Comparison of the averaged mass fractions in the two-zone model with the
 solar system abundances~\cite{Anders1989} (indicated by dots).}
\label{fig:massfrac_obs}

\end{figure}

%Proto type ----------------------------------------------------------------
Figure~\ref{fig:massfrac_obs} shows the abundances 
averaged between high- and low-density region using Eq.~\eqref{eq:Yi_ave}
compared with the solar system abundances~\cite{Anders1989}.
For $\eta^{}_{high}\simeq 10^{-4}_{}$, abundance productions of $120<A<180$ are
comparable to the solar values. For $\eta^{}_{high}\simeq10^{-3}_{}$,
those of $50<A<100$ have been synthesized well.
In the case of  $\eta^{}_{high}=5.3\times10^{-4}$, 
there are outstanding two peaks; one is around $A=56~(N=28)$ and the other can
be found around $A=140$.
Abundance patterns are very different from that of the solar system,
because IBBN  occurs under the condition of significant abundances of
both neutrons and protons.
%---------------------------------------------------------------------------

\begin{table*}[tb]
\caption{\label{tab:abundance_4463} 
Mass fractions of light elements for follow cases :
 $\eta^{}_{high}\simeq10^{-3}_{}$, $\eta_{high}=5\times10^{-4}$, 
 $\eta^{}_{high}\simeq10^{-4}$, and $\eta^{}_{high}=10^{-5}$. $t^{}_{fin}$ and $T^{}_{fin}$ is the time
 and temperature at the final stage of the calculations.}
{%\small
\begin{ruledtabular}
\begin{tabular}{ccccccc}
%\hline\hline
$f^{}_v, R $
& \multicolumn{3}{c}{$2.1\times10^{-8}_{}, 1.8\times10^{6}_{}$}
& \multicolumn{3}{c}{$7.0\times10^{-8}_{}, 9.3\times10^{5}_{}$}
%& \multicolumn{3}{c}{$1.0\times10^{-7}_{}, 1.7\times10^{5}_{}$}
 \\
%\hline 
$( \eta^{}_{high}, \eta^{}_{low} )$  
& \multicolumn{3}{c}{($1.06\times10^{-3}_{}$, $5.88\times10^{-10}_{}$) }
& \multicolumn{3}{c}{($5.33\times10^{-4}_{}$, $5.73\times10^{-10}_{}$) }
%& \multicolumn{3}{c}{$1.02\times10^{-4}_{}$, $6.00\times10^{-10}_{}$}
 \\
\hline
$\left( t^{}_{fin}, T^{}_{fin} \right)$ & 
\multicolumn{3}{c}{$1.0\times10^{5}_{}$sec,~$4.2\times10^{7}_{}$ K} &
\multicolumn{3}{c}{$1.1\times10^{5}_{}$sec,~$4.9\times10^{7}_{}$ K} 
%\multicolumn{3}{c}{$1.2\times10^{5}_{}$sec,~$4.3\times10^{7}_{}$ K} 
\\
\hline%\hline
elements & high & low & average & high & low & average % &  high & low & average 
\\
\hline 
p   & $0.586$ & $0.753$ & $0.746$ & $0.598$  &  $0.753$ &  $0.743$  \\
D     & $1.76\times10^{-21}_{}$ & $4.48\times10^{-5}_{}$ &  $4.32\times10^{-5}_{}$ 
      & $4.14\times10^{-21}_{}$ & $4.67\times10^{-5}_{}$ &  $4.38\times10^{-5}_{}$ 
%      & $6.84\times10^{-22}_{}$ & $4.34\times10^{-5}_{}$ &  $4.27\times10^{-5}_{}$
\\
%He3 &   $2.9\times10^{-14}_{}$ &  $2.2\times10^{-5}_{}$ & $9.3\times10^{-6}_{}$ \\
${}^4_{}$He 
& $0.413$ &   $0.247$  &   $0.253$
& $0.402$ & $0.247$ & $0.257$ 
\\
%& $0.362$ & $0.248$ & $0.249$ \\
${}^7_{}$Li &
  $1.63\times10^{-13}_{}$ &  $1.79\times10^{-9}_{}$ & $1.72\times10^{-9}_{}$ 
& $3.43\times10^{-13}_{}$ &  $1.70\times10^{-9}_{}$ & $1.59\times10^{-9}_{}$ 
%&   $7.42\times10^{-13}_{}$ &  $1.87\times10^{-9}_{}$ & $1.70\times10^{-9}_{}$ 
\\
%\hline\hline
\end{tabular}
{(a) For cases of $\eta^{}_{high}=10^{-3}$ and $\eta^{}_{high}=5\times10^{-4}_{}$.}

% table I(b)
\begin{tabular}{ccccccc}
%\hline\hline
$f^{}_v, R $ 
%& \multicolumn{3}{c}{$2.1\times10^{-8}_{}, 1.8\times10^{6}_{}$}
%& \multicolumn{3}{c}{$7.0\times10^{-8}_{}, 9.3\times10^{5}_{}$}
& \multicolumn{3}{c}{$1.0\times10^{-7}_{}, 1.7\times10^{5}_{}$}
& \multicolumn{3}{c}{$7.5\times10^{-6}_{}, 1.9\times10^{4}_{}$}
 \\
%\hline 
$(\eta^{}_{high}, \eta^{}_{low} )$  
%& \multicolumn{3}{c}{$1.06\times10^{-3}_{}$, $5.88\times10^{-10}_{}$ }
%& \multicolumn{3}{c}{$5.33\times10^{-4}_{}$, $5.73\times10^{-10}_{}$ }
& \multicolumn{3}{c}{($1.02\times10^{-4}_{}$, $6.00\times10^{-10}_{}$)}
& \multicolumn{3}{c}{($1.02\times10^{-5}_{}$, $5.34\times10^{-10}_{}$)}
 \\
\hline
$\left( t^{}_{fin}, T^{}_{fin} \right)$ & 
%\multicolumn{3}{c}{$1.0\times10^{5}_{}$sec,~$4.2\times10^{7}_{}$ K} &
%\multicolumn{3}{c}{$1.1\times10^{5}_{}$sec,~$4.9\times10^{7}_{}$ K} &
\multicolumn{3}{c}{$1.2\times10^{5}_{}$sec,~$4.3\times10^{7}_{}$ K} &
\multicolumn{3}{c}{$1.2\times10^{5}_{}$sec,~$4.5\times10^{7}_{}$ K} 
\\
\hline%\hline
elements & high & low & average & high & low & average% &  high & low & average 
\\
\hline 
p   & $0.638$  &  $0.752$ &  $0.750$ &  $0.670$ &   $0.753$ &   $0.743$ \\
D %    & $1.76\times10^{-21}_{}$ & $4.48\times10^{-5}_{}$ &  $4.32\times10^{-5}_{}$ 
  %    & $4.14\times10^{-21}_{}$ & $4.67\times10^{-5}_{}$ &  $4.38\times10^{-5}_{}$ 
      & $6.84\times10^{-22}_{}$ & $4.34\times10^{-5}_{}$ &  $4.27\times10^{-5}_{}$
      & $1.12\times10^{-22}_{}$ & $5.19\times10^{-5}_{}$ &  $4.55\times10^{-5}_{}$
\\
%He3 &   $2.9\times10^{-14}_{}$ &  $2.2\times10^{-5}_{}$ & $9.3\times10^{-6}_{}$ \\
${}^4_{}$He 
%& $0.413$ &   $0.247$  &   $0.253$
%& $0.402$ & $0.247$ & $0.257$   
& $0.362$ & $0.248$ & $0.249$ 
& $0.330$ & $0.246$ & $0.257$
\\
${}^7_{}$Li %&
%  $1.63\times10^{-13}_{}$ &  $1.79\times10^{-9}_{}$ & $1.72\times10^{-9}_{}$ 
%&  $3.43\times10^{-13}_{}$ &  $1.70\times10^{-9}_{}$ & $1.59\times10^{-9}_{}$ 
&   $7.42\times10^{-13}_{}$ &  $1.87\times10^{-9}_{}$ & $1.70\times10^{-9}_{}$ 
& $6.73\times10^{-8}_{}$  &  $1.47\times10^{-9}_{}$ & $9.63\times10^{-9}_{}$ 
\\
%\hline\hline
\end{tabular}
{(b) For cases of $\eta^{}_{high}=10^{-4}$ and $\eta^{}_{high}=10^{-5}_{}$.}
\end{ruledtabular}
}
\end{table*}

\begin{table*}[tb]
\caption{\label{tab:abundance_4463_2}Mass fractions of heavy elements
 $( A>7 )$ for three cases of
 $\eta^{}_{high}\simeq10^{-3}_{}, \eta_{high}=5.33\times10^{-4}$, 
and $\eta^{}_{high}\simeq10^{-4}$.}
\begin{ruledtabular}
\begin{tabular}{ccccccccc}
%\hline\hline
%$~f^{}_v, R~$
\multicolumn{3}{c}{$f^{}_v=2.1\times10^{-8}_{}, R=1.8\times10^{6}_{}$}
& \multicolumn{3}{c}{$f^{}_v=7.0\times10^{-7}_{}, R=9.3\times10^{5}_{}$}
& \multicolumn{3}{c}{$f^{}_v=5.7\times10^{-7}_{}, R=1.9\times10^{5}_{}$}
 \\
%\hline
%$~\eta^{}_{high}$, $\eta^{}_{low}~$
 \multicolumn{3}{c}{$\left( \eta^{}_{high}=1.06\times10^{-3}_{} \right)$}
& \multicolumn{3}{c}{$\left( \eta^{}_{high}=5.33\times10^{-4}_{} \right)$}
& \multicolumn{3}{c}{$\left( \eta^{}_{high}=1.02\times10^{-4}_{} \right)$}
 \\
\hline
%$~t_{fin}, T_{fin}~$ &   
%\multicolumn{3}{c|}{~$t_{fin}=1.0\times10^{5}_{}$~sec,~$T_{fin}=4.2\times10^{7}_{}$~K~} &
%\multicolumn{3}{c|}{~$t_{fin}=1.1\times10^{5}_{}$~sec,~$T_{fin}=4.9\times10^{7}_{}$~K~} &
%\multicolumn{3}{c}{~$t_{fin}=1.2\times10^{5}_{}$~sec,~$T_{fin}=4.3\times10^{7}_{}$~K~} \\
%\hline%\hline
element & high & average & 
element & high & average & 
element & high & average  \\
\hline\hline
 ${}^{56}_{}$Ni & $1.25\times10^{-4}_{}$ & $4.55\times10^{-6}_{}$ &
 ${}^{142}$Nd   & $1.99\times10^{-5}_{}$ & $1.22\times10^{-6}_{}$ &
%  ${}^{12}_{}$C  & $7.67\times10^{-8}_{}$ & $1.28\times10^{-9}_{}$ 
  ${}^{145}_{}$Nd& $3.69\times10^{-7}_{}$ & $3.61\times10^{-8}_{}$ 
\\
 ${}^{57}_{}$Co & $1.59\times10^{-5}_{}$ & $5.79\times10^{-7}_{}$ &
 ${}^{56}_{}$Ni & $1.37\times10^{-5}_{}$ & $8.35\times10^{-7}_{}$ &
  ${}^{40}_{}$Ca & $2.71\times10^{-7}_{}$ & $2.65\times10^{-8}_{}$
%  ${}^{13}_{}$C  & $6.43\times10^{-8}_{}$ & $1.07\times10^{-9}_{}$
\\
 ${}^{86}_{}$Sr & $1.06\times10^{-5}_{}$ & $3.86\times10^{-7}_{}$ &
 ${}^{146}$Sm   & $1.03\times10^{-5}_{}$ & $6.30\times10^{-7}_{}$ &
  ${}^{52}_{}$Mn & $2.42\times10^{-7}_{}$ & $2.36\times10^{-8}_{}$ 
%  ${}^{14}_{}$N  & $8.74\times10^{-8}_{}$ & $1.46\times10^{-9}_{}$ 
\\
 ${}^{87}_{}$Sr & $9.77\times10^{-6}_{}$ & $3.56\times10^{-7}_{}$ &
 ${}^{145}_{}$Pm& $8.91\times10^{-6}_{}$ & $5.44\times10^{-7}_{}$ &
  ${}^{155}_{}$Eu& $2.37\times10^{-7}_{}$ & $2.32\times10^{-8}_{}$
%  ${}^{16}_{}$O  & $6.30\times10^{-8}_{}$ & $1.05\times10^{-9}_{}$ 
\\
 ${}^{74}_{}$Se & $9.75\times10^{-6}_{}$ & $3.55\times10^{-7}_{}$ &
 ${}^{148}_{}$Sm& $8.25\times10^{-6}_{}$ & $5.05\times10^{-7}_{}$ &
  ${}^{140}_{}$Ce& $1.93\times10^{-7}_{}$ & $1.89\times10^{-8}_{}$
%  ${}^{40}_{}$Ca & $2.71\times10^{-7}_{}$ & $4.52\times10^{-9}_{}$
\\
 ${}^{75}_{}$Se & $2.93\times10^{-6}_{}$ & $1.07\times10^{-7}_{}$ &
 ${}^{147}_{}$Pm& $6.62\times10^{-6}_{}$ & $4.05\times10^{-7}_{}$ &
  ${}^{51}_{}$Cr & $1.55\times10^{-7}_{}$ & $1.51\times10^{-8}_{}$
%  ${}^{8}_{}$Be & $2.49\times10^{-9}_{}$ & $4.16\times10^{-11}_{}$
\\
 ${}^{74}_{}$Sr & $9.17\times10^{-6}_{}$ & $3.34\times10^{-7}_{}$ &
 ${}^{144}_{}$Sm & $5.24\times10^{-6}_{}$ & $3.20\times10^{-7}_{}$ &
  ${}^{142}_{}$Ce& $1.11\times10^{-7}_{}$ & $1.09\times10^{-8}_{}$
%  ${}^{48}_{}$Cr & $8.56\times10^{-8}_{}$ & $1.43\times10^{-9}_{}$
\\
 ${}^{82}_{}$Kr & $8.91\times10^{-6}_{}$ & $3.25\times10^{-7}_{}$ &
 ${}^{143}_{}$Pm & $4.15\times10^{-6}_{}$ & $2.54\times10^{-7}_{}$ &
  ${}^{56}_{}$Ni & $1.10\times10^{-7}_{}$ & $1.08\times10^{-8}_{}$
%  ${}^{44}_{}$Ti & $7.78\times10^{-9}_{}$ & $1.30\times10^{-10}_{}$
\\
 ${}^{81}_{}$Kr & $7.80\times10^{-6}_{}$  & $2.84\times10^{-7}_{}$ &
 ${}^{147}_{}$Sm & $3.99\times10^{-6}_{}$  & $2.44\times10^{-7}_{}$ &
 ${}^{146}_{}$Nd & $1.05\times10^{-7}_{}$  & $1.03\times10^{-8}_{}$
% ${}^{134}_{}$Xe & $2.24\times10^{-8}_{}$  & $3.74\times10^{-10}_{}$
\\
 ${}^{72}_{}$Ge  &  $7.67\times10^{-6}_{}$ & $2.80\times10^{-7}_{}$ &
 ${}^{144}_{}$Pm & $3.66\times10^{-6}_{}$  & $2.24\times10^{-7}_{}$ &
  ${}^{156}_{}$Eu &  $9.44\times10^{-8}_{}$ & $9.22\times10^{-9}_{}$
%  ${}^{136}_{}$Xe &  $2.77\times10^{-8}_{}$ & $4.63\times10^{-10}_{}$
\\ %${}^{}_{}$
 ${}^{78}_{}$Kr  &  $7.60\times10^{-6}_{}$ & $2.77\times10^{-7}_{}$ &
 ${}^{146}_{}$Pm & $3.46\times10^{-6}_{}$  & $2.12\times10^{-7}_{}$ &
  ${}^{148}_{}$Nd &  $9.36\times10^{-8}_{}$ & $9.15\times10^{-9}_{}$
%  ${}^{137}_{}$Cs &  $5.56\times10^{-8}_{}$ & $9.29\times10^{-10}_{}$
\\
 ${}^{80}_{}$Kr  &  $7.06\times10^{-6}_{}$ & $2.57\times10^{-7}_{}$ &
 ${}^{143}_{}$Nd & $2.84\times10^{-6}_{}$  & $1.74\times10^{-7}_{}$ &
   ${}^{52}_{}$Fe  & $8.97\times10^{-8}_{}$ &  $8.77\times10^{-9}_{}$
%  ${}^{17}_{}$O   &  $1.88\times10^{-9}_{}$ & $3.14\times10^{-11}_{}$
%  ${}^{136}_{}$Cs &  $3.95\times10^{-9}_{}$ & $6.59\times10^{-11}_{}$
\\
  ${}^{83}_{}$Kr  & $6.25\times10^{-6}_{}$ &  $2.28\times10^{-7}_{}$ &
 ${}^{145}_{}$Sm & $2.67\times10^{-6}_{}$  & $1.63\times10^{-7}_{}$ &
   ${}^{161}_{}$Tb & $8.96\times10^{-8}_{}$ &  $8.52\times10^{-9}_{}$
%   ${}^{114}_{}$Cd & $6.43\times10^{-9}_{}$ &  $1.08\times10^{-10}_{}$
\\
  ${}^{73}_{}$Ge  & $6.14\times10^{-6}_{}$ &  $2.24\times10^{-7}_{}$ &
 ${}^{144}_{}$Nd & $2.25\times10^{-6}_{}$  & $1.37\times10^{-7}_{}$ &
   ${}^{139}_{}$La & $8.80\times10^{-8}_{}$ &  $8.60\times10^{-9}_{}$
\\
  ${}^{76}_{}$Se   &  $5.93\times10^{-6}_{}$ &  $2.16\times10^{-7}_{}$ &
 ${}^{149}_{}$Sm & $1.76\times10^{-6}_{}$  & $1.07\times10^{-7}_{}$ &
   ${}^{14}_{}$N     &  $8.74\times10^{-8}_{}$ &  $8.54\times10^{-9}_{}$
\\
  ${}^{79}_{}$Br  &  $5.90\times10^{-6}_{}$ &  $2.15\times10^{-7}_{}$ &
 ${}^{148}_{}$Pm & $1.16\times10^{-6}_{}$  & $7.09\times10^{-7}_{}$ &
   ${}^{48}_{}$Cr   &  $8.56\times10^{-8}_{}$ &  $8.39\times10^{-9}_{}$
\\
  ${}^{77}_{}$Se  &  $5.35\times10^{-6}_{}$ &  $1.95\times10^{-7}_{}$ &
  ${}^{150}_{}$Sm & $9.88\times10^{-7}_{}$  &  $6.04\times10^{-7}_{}$ &
   ${}^{138}_{}$Ba &  $7.96\times10^{-8}_{}$ &  $7.77\times10^{-9}_{}$
\\
  ${}^{89}_{}$Y   &  $4.76\times10^{-6}_{}$ &  $1.73\times10^{-7}_{}$ &
   ${}^{57}_{}$Ni &  $7.52\times10^{-7}_{}$  &  $4.60\times10^{-7}_{}$ &
   ${}^{12}_{}$C   &  $7.67\times10^{-8}_{}$ &  $7.50\times10^{-9}_{}$
\\
  ${}^{90}_{}$Zr  &  $4.41\times10^{-6}_{}$ &  $1.61\times10^{-7}_{}$ &
  ${}^{108}_{}$Cd &  $5.92\times10^{-7}_{}$  &  $3.62\times10^{-7}_{}$ &
   ${}^{162}_{}$Dy &  $6.84\times10^{-8}_{}$ &  $6.68\times10^{-9}_{}$ 
\\
  ${}^{85}_{}$Rb  &  $4.32\times10^{-6}_{}$ &  $1.58\times10^{-7}_{}$ &
  ${}^{151}_{}$Eu &  $5.30\times10^{-7}_{}$ &  $3.24\times10^{-7}_{}$ &
   ${}^{13}_{}$C  &  $6.43\times10^{-8}_{}$ &  $6.28\times10^{-9}_{}$ 
\\
  ${}^{83}_{}$Rb  &  $4.08\times10^{-6}_{}$ &  $1.49\times10^{-7}_{}$ &
  ${}^{153}_{}$Eu &  $5.23\times10^{-7}_{}$ &  $3.20\times10^{-8}_{}$ &
  ${}^{16}_{}$O   &  $6.30\times10^{-8}_{}$ &  $6.16\times10^{-9}_{}$ 
\\
  ${}^{88}_{}$Y   &  $3.85\times10^{-6}_{}$ &  $1.40\times10^{-7}_{}$ &
  ${}^{110}_{}$Cd &  $4.70\times10^{-7}_{}$ &  $2.88\times10^{-8}_{}$ &
  ${}^{158}_{}$Gd &  $5.85\times10^{-8}_{}$ &  $5.71\times10^{-9}_{}$
\\
  ${}^{88}_{}$Zr  & $3.55\times10^{-6}_{}$  &  $1.29\times10^{-7}_{}$ &
  ${}^{149}_{}$Eu & $3.79\times10^{-7}_{}$  &  $2.32\times10^{-8}_{}$ &
  ${}^{137}_{}$Cs & $5.56\times10^{-8}_{}$  &  $5.43\times10^{-9}_{}$
\\
  ${}^{73}_{}$As  & $3.52\times10^{-6}_{}$  &  $1.28\times10^{-7}_{}$ &
  ${}^{152}_{}$Eu & $3.45\times10^{-7}_{}$  &  $2.11\times10^{-8}_{}$ &
  ${}^{147}_{}$Nd & $3.96\times10^{-8}_{}$  &  $3.87\times10^{-10}_{}$
\\
  ${}^{71}_{}$Ga  & $3.40\times10^{-6}_{}$  &  $1.23\times10^{-7}_{}$ &
  ${}^{140}_{}$Ce & $3.48\times10^{-7}_{}$  &  $2.12\times10^{-8}_{}$ &
  ${}^{165}_{}$Ho & $3.77\times10^{-8}_{}$  &  $3.68\times10^{-9}_{}$ 
\\
  ${}^{75}_{}$Se  & $2.93\times10^{-6}_{}$  &  $1.07\times10^{-7}_{}$ &
  ${}^{150}_{}$Eu & $2.90\times10^{-7}_{}$  &  $1.77\times10^{-8}_{}$ &
  ${}^{143}_{}$Pr & $3.11\times10^{-8}_{}$  &  $3.04\times10^{-9}_{}$ 
\\
  ${}^{91}_{}$Nb  & $2.90\times10^{-6}_{}$  &  $1.06\times10^{-7}_{}$ &
  ${}^{106}_{}$Cd & $2.89\times10^{-7}_{}$  &  $1.77\times10^{-8}_{}$ &
  ${}^{141}_{}$Ce & $3.00\times10^{-8}_{}$  &  $2.93\times10^{-10}_{}$ 
\\
\hline
$\displaystyle\sum_{A>7}{X(A)}$ &   $3.10\times10^{-4}_{}$
  & $1.13\times10^{-5}_{}$ & %Total $(A>7)$ 
$\displaystyle\sum_{A>7}{X(A)}$ &   $1.78\times10^{-4}_{}$
  & $7.20\times10^{-6}_{}$ & %Total $(A>7)$ 
  $\displaystyle \sum_{A>7}{X(A)}$ & $3.85\times10^{-6}_{}$  & $3.57\times10^{-7}_{}$ \\
%\hline\hline
\end{tabular}
\end{ruledtabular}
\end{table*}

\begin{figure}[t]
\begin{center}
 \includegraphics[width=.70\linewidth,keepaspectratio]{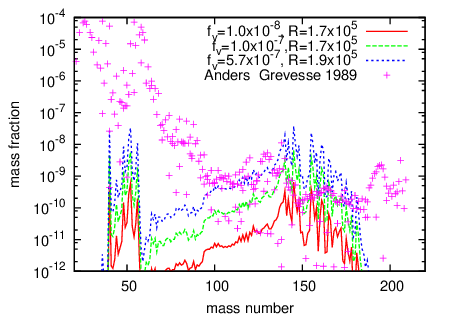}
\end{center}
\caption{Same as Fig.~\ref{fig:massfrac_obs}, but $\eta^{}_{high}$ is
fixed as $10^{-4}_{}$.}
\label{fig:massfrac_1e-4}
\end{figure}

% Present value of Hubble length is $c/H^{}_0\sim 1.28\times 10^{26}_{}$~m.

\section{Summary and Discussion}
\label{rec:summary}

We have investigated the consistency between inhomogeneous BBN  and the
observation of ${}^4_{}$He and D/H abundances under the standard
cosmological model having $\eta$ determined by WMAP.
We have adopted the two-zone model, where the universe has the high- and
low- baryon density regions at the BBN epoch.

First, we have calculated the light element nucleosynthesis 
using the BBN code having 24 nuclei for the high- and low-density regions.
We have assumed that the diffusion effect is negligible. 
There are significant differences for the time evolution of the
light element between the high- and low-density regions;
In the high-density region, the nucleosynthesis begins faster and
${}^4_{}$He is more abundant than that in the low density region as
shown in Figure~\ref{fig:MFNHSfig4}. 
From ${}^4_{}$He and D/H observations, we can put severe constraint on two
parameters of the two-zone model: the volume fraction $f^{}_v$ of
the high-density 
region and the density ratio $R$ between the two regions, 
where we have assumed that abundances in the two regions are mixed homogeneously.
%Furthermore, we also obtain the upper limit of CNO elements : $X(A>7) < 10^{-7}_{}$.

Second, using the allowed parameters constrained from the light element observations,
we calculate the nucleosynthesis that includes 4463 nuclei in the high-density
regions. Qualitatively, results of nucleosynthesis are the same as those in
Ref.~\cite{Matsuura2005}.
%%%%%
In the present results,
we showed that $p$- and $r$-elements are synthesized
simultaneously at high-density region with $\eta_{high}\simeq 10^{-4}$.d
Such a curious site of the nucleosynthesis have never been known in previous
studies of nucleosynthesis.

%%%%%
As the results, we have obtained the average values of mass fractions
from the nucleosynthesis  in high-density and that in low density regions.
The total averaged mass fractions beyond the light elements $X(A>7)$ are
constrained to be $10^{-5}_{}$ (for $\eta^{}_{high}=10^{-3}_{}$) 
and $10^{-7}$ (for $\eta^{}_{high}=10^{-4}_{}$).
We find that the average mass fractions in IBBN amount to
as much as the solar system abundances.
As see from Fig.~\ref{fig:massfrac_obs}, there are over-produced elements
around $A=150$ (for $\eta^{}_{high}=10^{-4}$) and $A=80$ (for
$\eta_{high}=10^{-3}$).
It seems to be conflict with the chemical evolution of the universe.
However, we show only the results of the upper-bounds on $f_v-R$
diagram. Since $f_v$ and $R$ are free-parameters, over-production
can be avoided by the adjustment of $f_v$ and/or $R$.
Figure \ref{fig:massfrac_1e-4} illustrates the mass fraction in
$\eta^{}_{high}=1.0\times10^{-4}_{}$ with various $f_v-R$ sets.
It is shown that the abundance pattern can be lower than the solar
system abundance.
Although we showed here only the result of $\eta^{}_{high}=10^{-4}_{}$ case, it is
possible to avoid producing over-abundance in other parameters,
$\eta_{high}=10^{-3}$ and $\eta^{}_{high}=5\times10^{-4}_{}$.
If we put constraint on the $f_v-R$ plane from the heavy element
observations, the limit of those parameters should be tightly.

In our calculation, 
the radioactive nuclei are produced much in the high-density region.
Especially, we should note that ${}^{56_{}}$Ni decays into ${}^{56_{}}$Fe 
(${}^{56}_{}$Ni$\rightarrow$ ${}^{56}_{}$Co$\rightarrow$
${}^{56}_{}$Fe), where the
existence of ${}^{56}_{}$Fe surely affects the process of the formation of
the first generation stars.
Therefore, it may be also necessary for IBBN to be constrained from
the star formation scenarios, because opacity change due to IBBN will
affect them.

Recent observational signal of over-abundances of ${}^{4}_{}$He mass
fractions in
globular clusters could motivate the IBBN scenario toward the detailed
modeling. The over-abundances of ${}^{4}_{}$He are suggested to be 
in the range of $0.3-0.4$ where estimated from the H-R diagram of the
blue Main-Sequence of NGC2808 in Ref.\cite{Piotto2007}.
If the origin of ${}^{4}_{}$He in globular clusters is due to IBBN, 
$\eta$ must be greater than $10^{-4}_{}$ in some regions during the epoch of
BBN. Then, the averaging procedure could be constrained from the more
detailed observations of abundances.
Since the history of changes in abundances has been investigated in
detail through the chemical evolution of galaxies~\cite{Anderson2009}, further plausible
constrains on the averaging process should be studied in the next step.

In our study, we ignore the diffusion effects.
However, it is shown that the diffusion affects the primordial
nucleosynthesis significantly~\cite{IBBN1}. Matsuura et
al.~\cite{Matsuura2007} has estimated the size of the high-baryon
density island to be $10^3$~m -- $10^{15}$~m at the BBN epoch.
The upper bound is obtained from the maximum angular resolution of CMB
and the lower is from the analysis of comoving diffusion length of neutron
and proton given in Ref.~\cite{IBBN0}.
In our case, we can estimate the scale of the high-density and the
effects of the diffusion from $f^{}_v$.

The neutron diffusion effects can be discussed with use of the results
obtained in the previous section by comparing the scale of the
high-density region with the diffusion length.
The present value of the Hubble length is $H^{-1}_0 = 1.28\times 10^{26}_{}$~m.
We may estimate the scale of the high-density region from the Hubble
length $H^{-1}$ multiplied by $f^{1/3}_v$. From ranges of the volume
fraction adopted in \S \ref{sec:Result2}, $10^{-8}< f^{}_v < 10^{-5}$, 
we obtain the scale of the high-density regions at present epoch $d_0$ as 
$2.7\times10^{23}$~m $< d_0 < 2.7\times10^{24}_{}$~m.
We can estimate the scale at redshift $z$ from the relation $d(z) = d^{}_0 (1+z)^{-1}$.
As the result, we expect $d$ at BBN era $(z\sim10^{9}_{})$ as 
$2.7\times10^{14}_{}$~m $< d^{}_{BBN} <  2.7\times 10^{15}_{}$~m.
We can say that the nucleon diffusion effects would be neglected
because the diffusion length is much smaller than $d$.

On the other hand, the high-density region is expected to be smaller
than $10^{15}_{}$~m. 
It seems to be very bad that the upper bound of $d$ is larger
than the value as far as our two zone model is concerned.
However, the high-density island cannot be observed directly, since we assume
that the high- and low-density
regions become homogeneous after the nucleosynthesis.
 
Finally, distances between high density regions are difficult
to derive without specific models beyond the two-zone model.
We will plan to calculate the nucleosynthesis with the diffusion of
abundances and/or more plausible averaging process included.

% \begin{figure}[t]
% \begin{center}
%  \includegraphics[width=0.8\linewidth,keepaspectratio]{figs_RN/Yp_high_ngc2808.eps}
% \end{center}
% \caption{\label{fig:with_ngc2808} Illustration of the helium mass fraction in the high-density region
%  against $\eta^{}_{high}$. Squared marks are the calculated values.
% The shaded-area corresponds to the
%  $Y^{}_p$ estimated in NGC2808~\cite{Piotto2007}.
% }
% \end{figure}

\acknowledgements
%\section{Acknowledgements}
%We would like to thanks to Dr. Nishimura for useful discussions.
This work has been supported in part by a Grant-in-Aid for Scientific
Research (18540279, 19104006, 21540272) of the Ministry of Education,
Culture, Sports, Science and Technology of Japan, 
and in part by a grant for Basic Science Research Projects 
from the Sumitomo Foundation (No. 080933).

%%   sample
%
%\begin{figure}[htbp]
%\begin{center}
%\includegraphics[height=6cm,width=9cm,clip]{mt-sora}
%\caption{The solar abundance and big bang nucleosynthesis}
%\label{fig-mt-solar}
%\end{center}
%\end{figure}
%
%%

%\appendix
%\section{First Appendix} %Empty argument \section{} yields `Appendix'. 
%
%\section{Second Appendix}


\begin{thebibliography}{99}
%%%%%%%%%%%%%%%%%%%%%%%%%%%%%%%%%%%%%%%%%%%%%%%%%%%%%%%%%%%%%
% Some macros are available for the bibliography:
%  o for general use
%    \JL : general journals                 \andvol : Vol (Year) Page
%  o for individual journal 
%    \AJ   : Astrophys. J.           \NC         : Nuovo Cim.
%    \ANN  : Ann. of Phys.           \NPA, \NPB  : Nucl. Phys. [A,B]
%    \CMP  : Commun. Math. Phys.     \PLA, \PLB  : Phys. Lett. [A,B]
%    \IJMP : Int. J. Mod. Phys.      \PRA - \PRE : Phys. Rev. [A-E]     
%    \JHEP : J. High Energy Phys.    \PRL        : Phys. Rev. Lett.
%    \JMP  : J. Math. Phys.          \PRP        : Phys. Rep.
%    \JP   : J. of Phys.             \PTP        : Prog. Theor. Phys.     
%    \JPSJ : J. Phys. Soc. Jpn.      \PTPS       : Prog. Theor. Phys. Suppl.
% Usage:
%  \PRD{45,1990,345}          ==> Phys.~Rev.\ \textbf{D45} (1990), 345
%  \JL{Nature,418,2002,123}   ==> Nature \textbf{418} (2002), 123
%  \andvol{B123,1995,1020}    ==> \textbf{B123} (1995), 1020
%%%%%%%%%%%%%%%%%%%%%%%%%%%%%%%%%%%%%%%%%%%%%%%%%%%%%%%%%%%%%
  
  %\cite{affl-dine}
%\bibitem{affl-dine}
%I. Affleck, and M. Dine, Nucl. Phys. {\bf B249}, 361 (1985).
\bibitem{Iocco:2008va}
	G.~Steigman,
	%``Primordial Nucleosynthesis in the Precision Cosmology Era,''
	Ann.\ Rev.\ Nucl.\ Part.\ Sci.\  {\bf 57}, 463 (2007); \\
	F.~Iocco, G.~Mangano, G.~Miele, O.~Pisanti and P.~D.~Serpico,
	Phys.\ Rept.\  {\bf 472}, 1 (2009)
 \bibitem{Luridiana2003}
	 V.~Luridiana,A.~Peimbert, M.~Peimbert, \&  M.~Cervino, 
	 Astrophys.\ J.\ {\bf 592}, 846 (2003)
\bibitem{Izotov:2007ed}
  Y.~I.~Izotov, T.~X.~Thuan and G.~Stasinska,
  %``The primordial abundance of 4He: a self-consistent empirical analysis of
  %systematic effects in a large sample of low-metallicity HII regions,''
  Astrophys.\ J.\  {\bf 662}, 15 (2007)
\bibitem{OliveSkillman04}
	Olive \& Skillman, Astrophys. J., {\bf 617}, 29--40 (2004)
\bibitem{Kirkman2003}
  D.~Kirkman, D.~Tytler, N.~Suzuki, J.~M.~O'Meara and D.~Lubin,
%   ``The cosmological baryon density from the deuterium to hydrogen ratio
  %towards QSO absorption systems: D/H towards Q1243+3047,''
  Astrophys.\ J.\ Suppl.\  {\bf 149}, 1 (2003)
  [arXiv:astro-ph/0302006].
  %%CITATION = ASTRO-PH 0302006;%%
\bibitem{Pettini2008} M. Pettini, B.~J. Zych, M.~T. Murphy, A. Lewis,
	\& C. C. Steidel, Mon. Not. R. Astron. Soc. {\bf 391}, 1499, (2008)
 \bibitem{OMeara2006}
	 J.~M.~O'Meara, S.~Burles, J.~X.~Prochaska, G.~E.~Prochter, R.~A.~Bernstein and K.~M.~Burgess,
	 %``The Deuterium to Hydrogen Abundance Ratio Towards the QSO SDSS1558-0031,''
	 Astrophys.\ J.\  {\bf 649}, L61 (2006)
\bibitem{Alcock1987}
	C.~Alcock, G.M.~Fuller, and G.J.~Mathews, Astrophys. J. {\bf 320}, 439 (1987)
 \bibitem{TerasawaSato89}
	 N.~Terasawa and K.~Sato,
	 Phys.\ Rev.\  D {\bf 39}, 2893 (1989)
\bibitem{2zone}
	K.~Jedamzik, and J.B.~Rehm, Phys. Rev. {\bf D64}, 023510 (2001)[astro-ph/0101292];\\
	T.~Rauscher, H.~Applegate, J.~Cowan, F.~Thielmann, and M.~Wiescher, \Apj {\bf 429},
	499 (1994).
\bibitem{IBBN0}
	J. H. Applegate, C. J. Hogan, and R. J. Scherrer, Phys. Rev. {\bf D35}, 1151 (1987)
\bibitem{IBBN1}
	R. M.~Malaney and W. A.~Fowler, Astrophys. J {\bf 333}, 14 (1988);\\ 
	J. H. Applegate, C. J. Hogan, R. J. Scherrer, Astrophys. J. {\bf 329}, 572 (1988);\\
	N. Terasawa and K. Sato, Astrophys. J. {\bf 362}, L.47 (1990);\\
	D.~Thomas, D.~N.~Schramm, K.A.~Olive, G.~J.~Mathews, B.~S.~Meyer, and B.~D.~
	Fields, \Apj {\bf430}, 291 (1994);
 \bibitem{Jedamzik1994}
	K.~Jedamzik, G.~M.~Fuller, G.~J.~Mathews, and T.~Kajino, \Apj {\bf 422}, 423 (1994);
\bibitem{Matsuura:2004ss}
  S.~Matsuura, A.~D.~Dolgov, S.~Nagataki and K.~Sato,
  %``Affleck-Dine Baryogenesis and heavy elements production from Inhomogeneous
  %Big Bang Nucleosynthesis,''
  Prog.\ Theor.\ Phys.\  {\bf 112}, 971 (2004)
\bibitem{Fuller1988}
	G.~M.~Fuller, G.~J.~Mathews and C.~R.~Alcock, Phys.\ Rev.\  D {\bf 37}, 1380 (1988);\\
\bibitem{IBBN_QCD}
	H.~Kurki-Suonio and R.~A.~Matzner, Phys.Rev. {\bf D39}, 1046 (1989);  \\
	H.~Kurki-Suonio and R.~A.~Matzner, Phys.Rev. {\bf D42}, 1047 (1990); 
\bibitem{WMAP3}
	C.L. Bennett, {\it et al.}, \Apj Suppl. {\bf 148}, 1 (2003) \\
	D.~N.~Spergel {\it et al.},  Astrophys.\ J.\ Suppl.\  {\bf 170},
	377 (2007) \\
	J.~Dunkley {\it et al.}   Astrophys.\ J.\ Suppl.\  {\bf 180}, 306 (2009) 
\bibitem{WMAP5}
	  E.~Komatsu {\it et al.},
	%``Seven-Year Wilkinson Microwave Anisotropy Probe (WMAP) Observations:
	%Cosmological Interpretation,''
	arXiv:1001.4538 [astro-ph.CO].
\bibitem{Juarez2009} 
	Y. Juarez, R. Maiolino, R. Mujica, M. Pedani, 
	S. Marinoni, T. Nagao, A. Marconi, \& E. Oliva, Astron. \& Astrophys.,
	494, L25, (2009)
%\bibitem{Boksenberg:2003}
%	A.~Boksenberg, W.~L.~W.~Sargent and M.~Rauch,
	%``Properties of QSO Metal Line Absorption Systems at High Redshifts: Nature
	%and Evolution of the Absorbers and the Ionizing Radiation Background,''
%	arXiv:astro-ph/0307557.
\bibitem{Bedin2004}
	L. R. Bedin et al., Astrophys. J., {\bf 605}, L125 (2004); 
\bibitem{Piotto2007}
	G. Piotto et al., Astrophys. J., {\bf 661} L53, (2007)
\bibitem{Rauther2006}
	T.~Rauscher, Phys.\ Rev.\  D {\bf 75}, 068301 (2007)
\bibitem{Matsuura2005}
	S.~Matsuura, S.~I.~Fujimoto, S.~Nishimura, M.~A.~Hashimoto and K.~Sato,
  %``Heavy Element Production in Inhomogeneous Big Bang Nucleosynthesis,''
	Phys.\ Rev.\  D {\bf 72}, 123505 (2005)
\bibitem{Matsuura2007}
	S.~Matsuura, S.~I.~Fujimoto, M.~A.~Hashimoto and K.~Sato,
  %``Reply to 'Comment on 'Heavy element production in inhomogeneous big bang
  %nucleosynthesis'',''
	Phys.\ Rev.\  D {\bf 75}, 068302 (2007).
 \bibitem{jedam}
	 K.~Jedamzik [astro-ph/9911242].
\bibitem{fujimoto} 
	S. Fujimoto,M. Hashimoto, O. Koike,K. Arai, \& R. Matsuba, \Apj {\bf
	585}, 418 (2003),\\
	O. Koike, M. Hashimoto, R. Kuromizu, \& S. Fujimoto, \Apj {\bf603}, 592 (2004),\\
	S. Fujimoto, M. Hashimoto, K. Arai, \& R. Matsuba, \Apj,
	{\bf 614}, 847 (2004),\\
	S.~Nishimura, K.~Kotake, M.~Hashimoto, S.~Yamada, N.~Nishimura,
	S.~Fujimoto and K.~Sato, Astrophys.\ J.\  {\bf 642}, 410 (2006).
\bibitem{Hashimoto1985}
	M. Hashimoto \& K. Arai, Physics Reports of Kumamoto University,
	{\bf 7}, 47, (1985).
\bibitem{PDG2008}
	B.~Fields and S.~Sarkar,
	%``Big-bang nucleosynthesis (PDG mini-review),''
	arXiv:astro-ph/0601514.
\bibitem{Kawano}
	L. Kawano, FERMILAB-Pub-92/04-A  
\bibitem{Anders1989}
	E.~Anders and N.~Grevesse,
  %``Abundances Of The Elements: Meteroritic And Solar,''
	Geochim.\ Cosmochim.\ Acta {\bf 53}, 197 (1989).
\bibitem{NACRE}
	C.~Angulo, M.~Arnould, M.~Rayet, P.~Descouvemont, D.~Baye,
	C.~ Leclercq-Willain, A.~Coc, S.~Barhoumi, P.~Aguer,
	C.~Rolfs, et al., Nuclear Physics A 656, 3 (1999).
\bibitem{Hashimoto1995} 
		M. Hashimoto, Progress of Theoretical Physics, 94, 663, (1995).
\bibitem{Hagiwara:2002fs}
	K.~Hagiwara {\it et al.}  [Particle Data Group],
  %``Review of particle physics,''
	Phys.\ Rev.\  D {\bf 66}, 010001 (2002).
\bibitem{Wagoner1967}
	R.~V. Wagoner, W.~A. Fowler,  \& F. Hoyle, 
	Astrophys.\  J.\ , {\bf 148}, 3 (1967)
\bibitem{Anderson2009}
  M.~E.~Anderson, J.~N.~Bregman, S.~C.~Butler and C.~R.~Mullis,
  Astrophys.\ J.\  {\bf 698}, 317 (2009)
\bibitem{Moriya2010}
  T.~Moriya and T.~Shigeyama,
  %``Multiple Main Sequence of Globular Clusters as a Result of Inhomogeneous
  %Big Bang Nucleosynthesis,''
  Phys.\ Rev.\  D {\bf 81}, 043004 (2010)
\end{thebibliography}
\end{document}